\documentstyle[prl,aps,preprint,epsf]{revtex}
\begin{document}
\draft
\bibliographystyle{prsty} 

\title{Deformed Fermi Surface Theory of Magneto--Acoustic Anomaly 
in Modulated Quantum Hall Systems Near $\nu=1/2$} 
\author{Nataliya A. Zimbovskaya$^{*}$ and Joseph L. Birman$^{**}$}
\address{Department of Physics, The City College of CUNY, New York,
NY, 10031, USA} 
\date{Revised \today} 
\maketitle

\begin{abstract}
 We introduce a new generic model of a deformed Composite Fermion--Fermi 
Surface (CF--FS) for the Fractional Quantum Hall Effect near
$\nu=1/2$ in 
the presence of a periodic density modulation. Our model permits us to 
explain recent surface acoustic wave  observations 
of anisotropic anomalies [1,2] in 
sound velocity and attenuation -- appearance of peaks and anisotropy
-- which 
originate from contributions to the conductivity tensor due to regions 
of the CF--FS which are flattened by  the applied modulation.
The calculated magnetic field and wave 
vector dependence of the CF conductivity,  velocity shift and attenuation
agree with experiments.
              \end{abstract}

PACS numbers 71.10 Pm, 73.40 Hm, 73.20 Dx {}

The integer and fractional quantum Hall effects (IQHE and FQHE) 
continue to reveal new and unexpected physics in strongly correlated
2--dimensional electron systems [3]. Recently, particular attention has
been given to FQHE systems at and near half filling of the lowest
Landau level (LLL). According to the Halperin, Lee and Read
theory [4] at $\nu = 1/2$ each
electron is decorated by two quantum flux tubes, producing a new
fermionic quasiparticle, the composite fermion (CF). At $ T=0, $
CFs are distributed inside the Composite Fermion--Fermi Surface
(CF--FS), which is assumed to be a circle.  At $ \Delta \nu = 0$, the
two flux tubes attached to each electron give rise via the ``Chern
Simons'' mechanism [3] to an extra (``fictitious'') magnetic field
opposite to and exactly canceling the applied ${\bf B}$ field. 
 When $\nu = 1/2 \pm \Delta \nu$ $(\Delta \nu \neq 0)$ the
Chern--Simons field does not cancel the applied field and the CF's move
in a non--zero magnetic field ${\bf B}_{eff} \; ({\bf B}_{eff} = B - 4
\pi \hbar c n/e ,$ where $  n $  is the electron density) which is
proportional
to $\Delta \nu $. In order to test the predictions of this theory it is
necessary to study the motion of the carriers in this field. A
sensitive tool for this purpose is surface acoustic wave (SAW)
propagation which gives quantitative information about the carriers
[5]. 

Recently [1,2] anomalous behavior was observed for the SAW velocity
and attenuation near filling factor $\nu = 1/2$ when a periodic
density modulation was applied. Measurements of the velocity shift
$\Delta s/s$ and the attenuation $\Gamma$ in the SAW response
orthogonal to the modulation direction showed an unexpected
effect. The minimum in $\Delta s/s$ at $\nu = 1/2$ which was observed
repeatedly in non--modulated systems [5], was converted to a large
maximum, when the modulation wave vector, and the magnitude of the
external field which produces the modulation, were above some critical
values. On further increase of the magnitude of the density modulation,
the peak in the velocity shift disappeared and was again replaced by a
minimum. For SAW propagation parallel to the direction of density
modulation, no such anomaly was found for the response of the electron
system.

The grating modulation will influence the CF system in two ways: 
through the direct effect of the modulating potential and through the
effect of the magnetic field $ \Delta B ({\bf r}) $ proportional to the
density modulation $ \Delta n ({\bf r}) \; \bigg(\Delta B ({\bf r}) = 4
\pi \hbar c \Delta n({\bf r})/e \bigg ). $ The latter was analysed
recently
[6]  under the conditions $ q << g; \; ql << 1\; ({\bf q, g} $ are
the SAW
and periodic density modulation wave vectors; $ l $ is the CF mean free
path).
It was shown that the corresponding component of the electron
conductivity $ (\sigma_{xx}, $ for the SAW propagating along the $
x $ axis
for ${\bf q} \perp {\bf g}$ had an additional term proportional to
$
(\Delta n/n)^2 \; (\Delta n $ is the amplitude of the density
modulation). Similar results under the same conditions 
were obtained in [7].

In this paper we will analyze the effect of the periodic density
modulation on the CF system under conditions of the experiment [1] 
$ (ql > 1, \; q \sim g). $ The starting point of the analysis is that
the periodic modulating field deforms the CF--FS analogous to  the
crystalline
field  in metals. The modulating potential wave vector {\bf g} in
this case replaces the reciprocal lattice vector.

We will show that a modulation-induced deformation of the originally
circular CF--FS can be at the origin of the observed transport
anomalies. We assume that exactly at $\nu = 1/2$ the CF-FS is a circle,
with radius $p_{F} = (4 \pi n \hbar^{2})^{1/2}. $ In the presence of the
grating modulation the CF--FS  circle is distorted and can be 
``flattened'' in the
neighborhood of  special points where the curvature vanishes. When $
ql > 1 $ such small, locally ``flat'' regions can under certain
conditions, play a disproportionately important role in determining the
magneto--conductivity response due to the unusually large density of
quasiparticle states there. The response is very sensitive to local
changes of the FS geometry: the flattening of the ''effective'' part of
the
CF--FS where the CF velocity vector {\bf v} and {\bf q} are nearly
transverse $ ({\bf q} \cdot {\bf v} \approx 0) $ can change the main
approximation to the CF conductivity whereas the nonuniformity of the $
B_{eff} $ determines corrections to it which are small when $ \Delta n/n
<< 1 $ [6,7]. Hence the modulation--induced deformation can be the most
important factor affecting the CF response functions at $ ql > 1. $ We
introduce a concrete model which permits us to obtain analytical
expressions for $\Delta s/s$ and $\Gamma$. Using appropriate parameters
we obtain semiquantative agreement with experiment. The model also
explains the orthogonality of response and predicts its wave-vector
dependence.

Our explicit deformed CF--FS model is new, to our knowledge.  A point of
contact between our work and that of [7] may be their assertion of
anisotropic resistivity due to the spatially averaged current and
electric field in the presence of periodically modulated quasiparticle
density [see eqn. (2) of ref. 7].  This assertion seems implicitly to
correspond to our deformed CF--FS; the two approaches would then be
equivalent when $\Delta n \ll n$. We conjecture below that the reason
for the reported disappearance of peaks at the highest modulation is
related to additional topological change in the CF--FS. 

As a first step, assume the periodic modulation in the 
y-direction introduces a single Fourier  component of potential $V_{g}$ 
into a  ``nearly--free'' particle CF model. The resulting dispersion 
relation is: 
   \begin{equation}
E({\bf p})= \frac{p_x^2}{2m^*} + 
\frac{p_y^{*2}}{2m^*} + 
\frac{(\hbar g)^2}{8m^*} - 
\sqrt{\left( \frac{\hbar g p_y^*}{2m^*} \right)^2 + V_{g}^{2}},
                         \end{equation}
with $p_{y}^{*}=p_{y}-\hbar g/2$, $m^{*}$ is the CF effective
mass. The curvature of the 2--D CF--FS can also be directly calculated as:
                 \begin{equation}
\kappa = \left [2v_{x} v_{y} \frac {\partial v_x}{\partial p_y}
- v_x^2 \frac {\partial v_y}{\partial p_y} - v_{y}^{2}
 \frac {\partial v_x}{\partial p_x} \right]  \left / v^{3}, \right.
                \end{equation} 
with $v=\sqrt{v_{x}^{2}+v_{y}^{2}}$. The
curvature $\kappa$ tends to zero when $ p_x \rightarrow \pm p_F \sqrt
{V_g/ E_F} $. The importance of this is that near to these points on the
CF--FS the CF velocities are nearly parallel to the $y$ direction. When
$ ql\gg 1$ these parts of the CF--FS make the major contribution to the
velocity shift $ \Delta s/s $ and attenuation $ \Gamma $ of the SAW
propagating in the $x$ direction. Near these zero curvature points we
will use asymptotic expressions for Eq.(1). Determining
$(p_{x0},p_{y0})$ by $p_{x0} = \eta p_{F} $, $ \displaystyle
{p_{y0} = p_{F} \left (1-\frac{1}{\sqrt{2}}\eta^{2} \right )} $, where
$\eta=\sqrt{V_{g}/E_{F}}$, $E_{F}=p_{F}^{2}/2m^{*}$, we can expand the
variable $p_{y}$ in powers of $(p_{x} - p_{x0})$, and keep the lowest
order terms in the expansion. We obtain: 
               \begin{equation}
p_{y}-p_{y0} = -\eta(p_{x}-p_{x0})-
\frac{2}{\eta^{4}}\frac{(p_{x}-p_{x0})^{3}}{p_{F}^{2}}.
                    \end{equation}
Near $p_{x0} $, where $(|p_{x} - p_{x0}|<\eta^{2}p_{F})$ the first
term on
the right side of Eq.(3) is small compared to the second one and can
be omitted.
Hence near $p_{x0} $ we have:
                       \begin{equation}
E({\bf p})=\frac{4}{\eta^{4}}\frac{p_{F}^{2}}{2m^{*}}
\left(\frac{p_{x}-p_{x0}}{p_{F}}\right)^{3}+\frac{p_{y}^{2}}{2m^{*}}.
                                   \end{equation}
The ''nearly free'' particle model can be used when the ratio $
V_g/E_F$ is very small. For larger $ V_g$ corresponding to $  \Delta
n/n$ of the order of a few percent (as in the experiment [1]) the
local flattening of the CF--FS can be more significant.
To analyze the contribution to the conductivity from these flattened
parts we  generalize Eq. (4) for $E({\bf p})$ and define our model
as:
          \begin{equation}
E({\bf p}) = \frac{p_{0}^{2}}{2m_{1}} \left |
\frac{p_{x}}{p_{0}} \right |^{\gamma} + \frac{p_{y}^{2}}{2m_{2}}, 
                                \end{equation}
where $p_{0}$ is a constant with the dimension of momentum, the
$m_{i}$ are effective masses, and $\gamma$ is a dimensionless
parameter which will determine the shape of the CF--FS . 
When $\gamma>2$ the 2--D CF--FS looks like an ellipse 
flattened near the vertices $(0, \pm p_{0})$. Near these points the 
curvature is:
                    \begin{equation}	
\kappa = -
\frac{\gamma(\gamma-1)}{2p_{0}\sqrt{m_{1}/m_{2}}}
\left |\frac{p_x}{p_0} \right |^{\gamma-2}
          \end{equation} 
and, $ \kappa \rightarrow  0$ at $p_{x} \rightarrow 0$. The CF--FS will be
the flatter
at $(0, \pm p_{0})$, the larger is the parameter $\gamma$.
A separate investigation is required to establish how $ \gamma $
depends on modulation magnitude $ V_g. $ Here we postulated Eq. (5)
as a natural generalization of Eq. (4) and we then derive the
resulting SAW response.A separate investigation is required to establish how $ \gamma $
depends on modulation magnitude $ V_g. $ Here we postulated Eq. (5)
as a natural generalization of Eq. (4) and we then derive the
resulting SAW response.

In a GaAs heterostructure with a 2-D electron gas subject to a 
travelling SAW, piezoelectric coupling 
produces a longitudinal electric field which interacts with the
electron gas. Taking the SAW wave vector as $(q,0,0)$ we obtain that
the resulting velocity shift $\Delta s/s$ and SAW attenuation rate
$\Gamma$ are given by the following expressions [8]
                        \begin{eqnarray}
\Delta s/s = [\alpha^2/2]
\Re (1 + i\sigma_{xx}/\sigma_{m} )^{-1}, \\   
\Gamma=-q(\alpha^{2})/2\Im(1+i\sigma_{xx}/\sigma_{m})^{-1}.
\end{eqnarray}
In these equations, $\omega=sq$ is the SAW frequency, $\alpha$ is the
piezoelectric coupling constant, $\sigma_{m}=\epsilon s/(2\pi)$ with
$\epsilon$ an effective dielectric constant of the medium, $\sigma_{xx}$
is the $xx$ component of the electronic conductivity tensor; real and
imaginary parts are indicated. In order to proceed we now need to
establish some preliminary results.  We use the semi--classical CF
theory [4] in which the CF quasiparticles have charge $e$, and finite
mass $ m^*. $ However, as described below, a particular variant of the
solution of the Boltzmann equation was needed for the present work. In
semiclassical CF theory the electron resistivity tensor $\rho$ at finite
$ {\bf q},\omega $ is the sum of a CF term and a term originating in the
magnetic field of the Chern-Simons (CS) vector potential. The CS part
has only off--diagonal elements, 
                   \begin{equation}
(\rho^{CS})_{xy}=-(\rho^{CS})_{yx} = 4 \pi \hbar/e^2.
                                               \end{equation}
In a strong magnetic field we have $\rho_{xy}\gg\rho_{xx}$, $\rho_{yy}$,
and hence we can use the approximation:
                         \begin{equation}
\sigma_{xx}({\bf q}) = \frac{e^4}{(4\pi\hbar)^2 }
\; \frac{\tilde \sigma_{xx}({\bf q})}{\tilde \sigma_{xx}({\bf q})
\tilde \sigma_{yy}({\bf q}) + \tilde \sigma_{xy}^2({\bf q})} \, ,
                       \end{equation}
where $\tilde{\sigma}=(\rho^{CF})^{-1}$ is the CF 
conductivity. 

To evaluate the CF conductivity
$\tilde{\sigma}_{\alpha\beta}({\bf q})$ so that we can pass
smoothly to the $B_{eff}\rightarrow 0$ limit for a flattened CF--FS, we
begin with the expression obtained from solution of the linearized
Boltzmann equation in the presence of the magnetic field, assuming a
relaxation time $\tau. $ This is:
                      \begin{eqnarray}
\tilde{\sigma}_{\alpha\beta}(\nu) =
 \frac{e^{2}m_{c}}{(2\pi\hbar)^{2}}\frac{1}{\Omega}
\int \limits_{0}^{2 \pi}d\psi \left \{ \exp \left
[-\frac{iq}{\Omega} \int \limits_{0}^{\psi}V_{x}
(\psi^{\prime\prime})d\psi^{\prime\prime}\right]v_{\alpha}(\psi) 
       \right.      \times \nonumber \\ 
\times     \left.
\int \limits_{-\infty}^{\psi} \exp
\left [\frac{iq}{\Omega} \int
\limits_{0}^{\psi'} v_{x}(\psi^{\prime})d\psi^{\prime}
+ \frac{1}{\Omega\tau}(\psi^{\prime} -\psi)\right ]
v_{\beta}(\psi^{\prime})d\psi^{\prime} \right \}.
                  \end{eqnarray}
Here $V_{\alpha,\beta} $ are the CF velocity components
$(\alpha,\beta=x,y)$; $\Omega=|e|B_{eff}/m_{c}c$ is their
cyclotron frequency; $\psi$ is the angular coordinate on the
CF cyclotron orbit, ($\psi=\Omega\theta$; $\theta$ is the
time of the CF motion along the cyclotron orbit). We have
taken $\omega\tau \ll 1$. We proceed [9] as follows. Express the
velocity components $ v_{\beta}(\psi^{\prime})$ as  Fourier series:
                        \begin{equation}
v_{\beta}(\psi^{\prime})=\sum_{k} v_{k\beta} \exp(ik\psi^{\prime}).
            \end{equation}
Introducing a new variable $\eta$:
                  \begin{equation}
\eta \equiv \left (\frac{1}{\tau} + ik\Omega+iq v_{x}(\psi)\right) 
{\theta}+iq\int \limits_{0}^{{\theta}}
[v_{x}(\psi+\Omega\theta^{\prime}) - v_{x}(\psi)]d\theta^{\prime};
\qquad         \theta = (\psi' - \psi)/\Omega
                        \end{equation}
and substituting (12) and (13) into (11) we obtain:
                  \begin{equation}
\tilde{\sigma}_{\alpha\beta}(\nu) = 
\frac{e^{2}m_c\tau}{(2\pi\hbar)^{2}}
\sum_{k} v_{k\beta} \int \limits_{-\infty}^{0}e^{\eta} d\eta
\int \limits_{0}^{2\pi}
\frac{v_{\alpha}(\psi)
\exp(ik\psi)d\psi}{1+ik\Omega\tau +
i q v_{x}(\psi+\tilde{\theta} (\eta)\Omega)\tau}\, .
                             \end{equation}
To proceed we can transform the integral over $\psi$ in (14) to an
integral over the CF--FS. Reexpressing the element of integration as
$ m_c d\psi = d \lambda /|v| \; \; (d\lambda $ is the element of
length along the Fermi Arc), and replacing $ m_c $ by a suitable
combination of $ m_1, m_2 $ of our model (5); e.g. for an ellipse $ m_c =
\sqrt {m_1 m_2}, $ we can now parameterize the dispersion equation
of our
model (5) as follows:
                   \begin{equation}
p_{x}=\pm p_{0}|\cos t|^{2/\gamma}; \qquad \qquad
p_{y}=p_{0}\sqrt{m_{2}/m_{1}} \, \sin t,
           \end{equation}
where $0\leq t\leq2\pi$, and the $+$ and $-$ signs are chosen
corresponding to normal domains of positive and negative values of the 
cosine. Where $ql \gg 1, $ the leading term in the resulting formula
originates from parts of the CF--FS where $v_{x}\approx 0$. Expanding
it in powers of $(ql)^{-1}$ and keeping the main term in the expansion
we obtain:
                           \begin{equation}
\tilde{\sigma}_{yy}(\nu) = 
\frac{b}{2} \frac{e^2 p_0}{4 \pi\hbar^2} \frac{l}{(ql)^\mu} 
(S_{+\mu}(\Omega\tau)+S_{-\mu}(\Omega\tau))
                                           \end{equation}
where: $\displaystyle {S_{\pm\mu}(\Omega\tau) = \int\limits_{-\infty}^{0}
e^\eta(1\mp i\Omega\tau(1\pm\eta\delta_{0}))^{\mu-1}d\eta} $ and
$\delta_{0}$ is a small dimensionless constant of the order of
$\omega\tau$.
Here for convenience we introduced $\mu=1/(\gamma-1)$ which is a 
dimensionless parameter $(0\not=\mu\leq 1)$, with $\mu=1$, or $\gamma=2$
corresponding to the case that the CF--FS is an ellipse.
In these variables, the CF mean-free-path $\ell$ is equal to:
                    \begin{equation}
\ell = \frac{\mu + 1}{2 \mu} \frac{p_0 \tau}{m_1}.
                                  \end{equation}
Passing to the limit $B_{eff}=0$ we have:
                                  \begin{equation}
\tilde{\sigma}_{yy} \left (\nu = \frac{1}{2} \right) =
\frac{be^2 p_0}{4\pi\hbar^2} \frac{\ell}{(q \ell)^\mu} \, . 
                                    \end{equation}
In this equation $ b =
4\mu^{2}/(\mu+1)\sqrt{m_{1}/m_{2}}[\sin(\pi\mu/2)]^{-1}$.
This expression eqn.(18)  predicts that  measuring the $q$-dependence of
the conductivity 
exactly at $B_{eff}=0 \;(\nu=1/2$) can give the deformation
parameter $\mu$. When the CF--FS is an undeformed circle
( $m_{1} = m_{2} = m^*) $ then $b=2$ and the result is identical to the 
corresponding result obtained in [4]. It is worth emphasizing that
when the flattening of the CF-FS is strong, with  $\gamma\gg 
1$, the quantity $\mu\approx 0$ and  the CF conductivity will be enhanced 
compared to the circular case, and it will be effectively independent
of q (See eqn.(16)). Independence of $q$ has been found experimentally
[1]. For small $\Omega\tau, \; (\Omega \tau \omega\tau < 1)$ one can
expand the functions $ S_{\pm \mu}(\Omega\tau) \; (\mu \neq 1) $ in
powers
of $\delta_{0}\Omega\tau$:
                         \begin{equation}
S_{\pm \mu}(\Omega\tau) = 
(1\mp\Omega\tau)^{\mu-1}\left[1+\sum_{r=1}^{\infty}
\frac{(1-\mu)(2-\mu)...(r-\mu)}{(1\mp
i\Omega\tau)^{r}}(i\delta_{0}\Omega\tau)^{r}\right].
                                                  \end{equation}
Keeping the terms larger than $(\Omega\tau)^3$ one has:
                         \begin{equation}
\tilde{\sigma}_{yy} = 
\tilde{\sigma}_{yy}  \left (\nu = \frac{1}{2} \right ) 
[1-a^{2}(\Omega\tau)^2 + i \xi \Omega\tau] .
                          \end{equation}
Here $a^{2}=((1-\mu)(2-\mu)/2)(1+2\delta_{0}^{2})$ and
$ \xi =(1-\mu)\delta_{0}$ are positive constants. For sufficiently small
values of the parameter $\mu$ (significant flattening of the effective
parts of the CF--FS) the constant $a^{2}$ is of the order of unity and
the constant $ \xi $ is small compared to unity, because of the small
factor $\delta_{0}$. Other components of the CF conductivity tensor can
be calculated similarly.

Substituting the results  into (10), we can obtain the expression
for the electron conductivity component $\sigma_{xx}$. Then using
(7),(8) we have:
                 \begin{equation}
\frac{\Delta s}{s} = \frac{\alpha^{2}}{2}
\frac{1+ \xi\Omega\tau\bar{\sigma}}{1+\bar{\sigma}^2}  
\left(1-\frac{2\xi\Omega\tau\bar{\sigma}}{1+\bar{\sigma}^{2}} - 
\frac{\bar{\sigma}^{2}}{1+\bar{\sigma}^{2}}
(2a^{2}-\xi^{2})(\Omega\tau)^{2}\right) ;
                                          \end{equation}
                         \begin{equation}
\Gamma = q\frac{\alpha^{2}}{2} 
\frac{\bar{\sigma}^{2}}{1+\bar{\sigma}^{2}}
\left(1-\frac{2 \xi\Omega\tau\bar{\sigma}}{1+\bar{\sigma}^{2}} 
-\frac{a^{2}\bar{\sigma}^{2}}{1+\bar{\sigma}^{2}}(\Omega\tau)^{2}\right) .
                                     \end{equation}
Here $\bar{\sigma}=\sigma_{xx}(\nu=1/2)/\sigma_{m}. $ Expression (21)
and (22) are the new results of our theory. They predict peaks both in
the SAW attenuation and velocity shift at $\nu=1/2; $ the peaks arise
due to distortion of the CF--FS in the presence of the density
modulation. When the CF--FS flattening is strong ($\mu\ll 1$)  the
magnitude of the peak of the velocity shift is practically independent
of the SAW wave vector $q$. Also these anomalies are not sensitive to
any relation between $q$ and the density modulation wave vector $g$. As
was observed repeatedly [1,2] the peaks appear when the magnitude of the
modulating potential and its wave vector are sufficiently large. These
quantities $V_{g}$ and $g$ determine the character and amount of
distortion of the CF--FS, by changing $ \gamma $ in our model.

We now suggest an explanation for the observed disappearance of the SAW
peak in $\Delta s/s$ when the magnitude of density modulation was at
highest measured values. In metals it is known [10,11] that external
factors, as well as changes in electron density can cause changes in FS
topology such as in the connectivity. These changes are sensitively
reflected in the response
functions. We suggest this can occur in the CF case. A topological
change of the CF--FS connectivity can be caused by increased magnitude
of modulating field and correspondingly increased quasiparticle density
modulation amplitude
$ {\Delta n} $. Changing the CF--FS connectivity can lead
to the disappearance of the flattening of the effective parts of the
CF--FS. In this case the anomalous maximum in the magnetic field
dependence of $\Delta s/s$ will be replaced by minimum. Thus assuming
the relevance of such a CF topological transition, we can explain
the
disappearance of the peak in the SAW velocity shift under increase of
the modulation strength. Additional experimental consequences of our
model and more details of the theory including the analysis of the
contributions to the 2DEG respose arising due to the additional
nonuniform magnetic field $ \Delta B({\bf r}) $ will be presented
elsewhere [12].

We again remark that our work is based on the charged CF picture
for FQHE, for example as derived at $\nu = 1/2 $ in ref. [4]. 
We followed previous work by assuming the CF--FS exists, as
supported also by a theoretical study [13]. The relevant magnetic
symmetry translation [14] which would replace Bloch wave vector
{\bf k} by a ''good'' quantum index for state labels and transport
theory, have not been used to our knowledge. An
alternate picture for the FQHE also derived from a Chern--Simons
approach, gives the quasiparticles at $\nu = 1/2$ as neutral
dipolar objects, with the Hall current being carried by a set of
collective magneto--plasmon oscillators. To our knowledge, a
magneto--transport theory based on this second picture does not
exist at present, so we are not able to compare our results with
any derived from that picture. 

We thank Dr.R.L.Willett, Dr.S.H.Simon for discussions, and
Dr.G.M.Zimbovsky for help with the manuscript. This work was supported
in part by a grant from the National Research Council COBASE Program. 

------------------------------------------------------------------

email is: *nzimbov@scisun.sci.ccny.cuny.edu and
**birman@scisun.sci.ccny.cuny.edu

1. R.L.Willet,  K.W.West and L.N.Pfeiffer, Phys. Rev. Lett., {\bf 78}, 
4478 (1997).

2. J.H.Smet, K. von Klitzing, D.Weiss and W.Wegscheider, Phys. Rev. Lett.,
{\bf 80}, 4538 (1998).

3. ''The Quantum Hall Effect'' ed R.E.Prange and S.M.Girvin (Springer,
NY 1987); ''Perspectives in Quantum Hall Effect''
ed. S. das Sarma and A.Pinczuk (J.Wiley, 1997).

4. B.I.Halperin, P.A.Lee and N.Read, Phys. Rev. B {\bf 47}, 7312 (1993); 
S.N.Simon and B.I.Halperin, Phys. Rev. B {\bf 48}, 17368 (1993);
A.Stern and B.I.Halperin, Phys. Rev. B {\bf 52}, 5840 (1995).

5. R.L.Willet. Advances of Physics, {\bf 46}, 447, (1997).

6. A.D.Mirlin, P.Wolfle, Y.Levinson and O.Entin-Wohlman,
cond-mat/9802140.

7. Felix v. Oppen, Ady Stern and B.I.Halperin,  Phys. Rev. Lett., 
{\bf 80}, 4494 (1998).

8. K.A.Inbergrigsten, J. Appl. Phys. {\bf 40}, 2681 (1969);
P.Bierbaum, Appl. Phys. Lett. {\bf 21}, 595 (1972).

9. N.A.Zimbovskaya, V.I.Okulov, A.Yu.Romanov and V.P.Silin,
Fiz. Met. Metalloved., {\bf 62}, 1095 (1986) (in Russian);
N.A.Zimbovskaya, Fiz. Nizk. Temp. {\bf 20}, 441 (1994); [Sov. Low
Temperature Physics, {\bf 20}, 324 (1994)].
N.A.Zimbovskaya, ''Local Geometry of the Fermi Surface 
and High Frequency Phenomena in Metals.'' ''Nauka'', Ekaterinburg, 1996
(in Russian).

10. I.M.Lifshitz, Zh. Eksp. Teor. Fiz. {\bf 38}, 1569 (1960)
[Sov. Phys JETP {\bf 11}, 1130 (1960)].

11. Ya.M.Blanter, M.I.Kaganov, A.V.Pantsulava and A.A.Varlamov, Physics
Reports, {\bf 245}, 159 (1994).

12. N.A.Zimbovskaya and J.L.Birman (in preparation).

13. P.Kopietz and G.E.Castilla, Phys. Rev. Lett. {\bf 78}, 314 (1997). 
 
14. P.B.Wiegmann and A.Zabrodin, Nuclear Physics B, {\bf 422}, 495
(1994).

\end{document}